\documentclass[aps,pra,twocolumn,superscriptaddress,nofootinbib,floatfix]{revtex4}

\usepackage{graphics}
\usepackage[usenames,dvipsnames]{color}
\usepackage{graphicx}
\usepackage{amssymb}
\usepackage{amsmath}
\usepackage{hyperref}

\usepackage{mathbbol}
\usepackage{wasysym} 
\usepackage{bbm}     
 
\usepackage{rotating}
\usepackage{dcolumn}
\usepackage{bm}
\usepackage{longtable}
\usepackage{multirow}
\usepackage{enumitem}

\newcommand{\ep}{\varepsilon}
\newcommand{\be}{\begin{equation}}
\newcommand{\ee}{\end{equation}}
\newcommand{\ba}{\begin{eqnarray}}
\newcommand{\ea}{\end{eqnarray}}
\newcommand{\ar}{\arrowvert} 
\newcommand{\ra}{\rangle} 
\newcommand{\la}{\langle}

\newcommand{\nts}{\negthickspace}
\newcommand{\nms}{\negmedspace}

\begin{document}
\title{Cubic wavefunction deformation of compressed atoms}
\author{Pedro Calvo Portela$^2$ and Felipe J. Llanes-Estrada} 
\affiliation{
Dept. F\'{\i}sica Te\'orica I and   
$^2$ F\'{\i}sica At\'omica, Molecular y Nuclear,\\
Universidad Complutense de Madrid, Facultad de Ciencias F\'{\i}sicas, Parque de las Ciencias 1, 28040 Madrid, Spain.}

\date{\today}

\begin{abstract}
\noindent
\renewcommand{\baselinestretch}{1.0}
We hypothesize that in a non-metallic crystalline structure under extreme pressures, atomic wavefunctions deform to adopt a reduced rotational symmetry consistent with minimizing interstitial space in the crystal. We exemplify with a simple numeric variational calculation that yields the energy cost of this deformation for Helium to 25\%.
Balancing this with the free energy gained by tighter packing we obtain the pressures required to effect such deformation.  
The consequent modification of the structure suggests a decrease in the resistance to tangential stress, and an associated decrease of the crystal's shear modulus. The atomic form factor is also modified. 
We also compare with neutron matter in the interior of compact stars. 
\large\normalsize
\end{abstract}

\maketitle

\section{Introduction}

When a solid is compressed, there is great energetic advantage in reducing the volume occupied per atom. Once an optimal close-packing structure is already in place, this can be achieved by a reduction of the interatomic spacing, or, what we study in this small note, a deformation of each of the atoms forming the solid. For a simple theoretical study Hydrogen is not the most appropriate study case because it forms covalent molecules, a four-body problem, so we settle to address Helium instead. It should be obvious to the reader that the ideas are extendible to other atomic systems, but the computations would be quite more challenging.

One can think of two obvious modes to deform the atomic wavefunction. The first is a simple compressional mode where the atom's radius is reduced but the lattice remains the same. The second is a mode in which the atom is deformed so as to abandon the spherical symmetry that it adopts in vacuum and instead makes a transition to an octahedral (cubic) symmetry consistent with the crystal axes, but maintaining the atomic volume constant. 
Because of the larger novelty of this concept, this is the mode that we will be addressing; it is plain that both should act on a real system under compression. 

Experimentally, Helium crystallizes at $T\simeq0\:\text{K}$ 
under a pressure of $2.5\:\text{MPa}$~\cite{prop_H_He} in an
hexagonal-compact (\textit{hcp}) structure, but presents also a face-centered cubic structure (\textit{fcc}) with a triple point fluid-\textit{fcc}-\textit{hcp} at $T\!=15\:\text{K}$ and $P\!=\!0.1\:\text{GPa}$~\cite{Dug_Sim}. 
Through X-ray diffraction studies, it has been determined that at $T\!=300\:\text{K}$ and high pressures between $15.6$ and $23.3\:\text{GPa}$, 
He is in the \textit{hcp} phase~\cite{xray_He}, and moreover the crystalline structure is deformed, diminishing the cell parameters due to the decrease in interatomic distance (the pure compressional mode). To our knowledge, the second mode of atomic deformation, breaking the central symmetry around the nucleus, has not been reported in the literature.

\section{Variational deformation energy\label{deformedenergy}}
\begin{figure}
\includegraphics[width=0.4\textwidth]{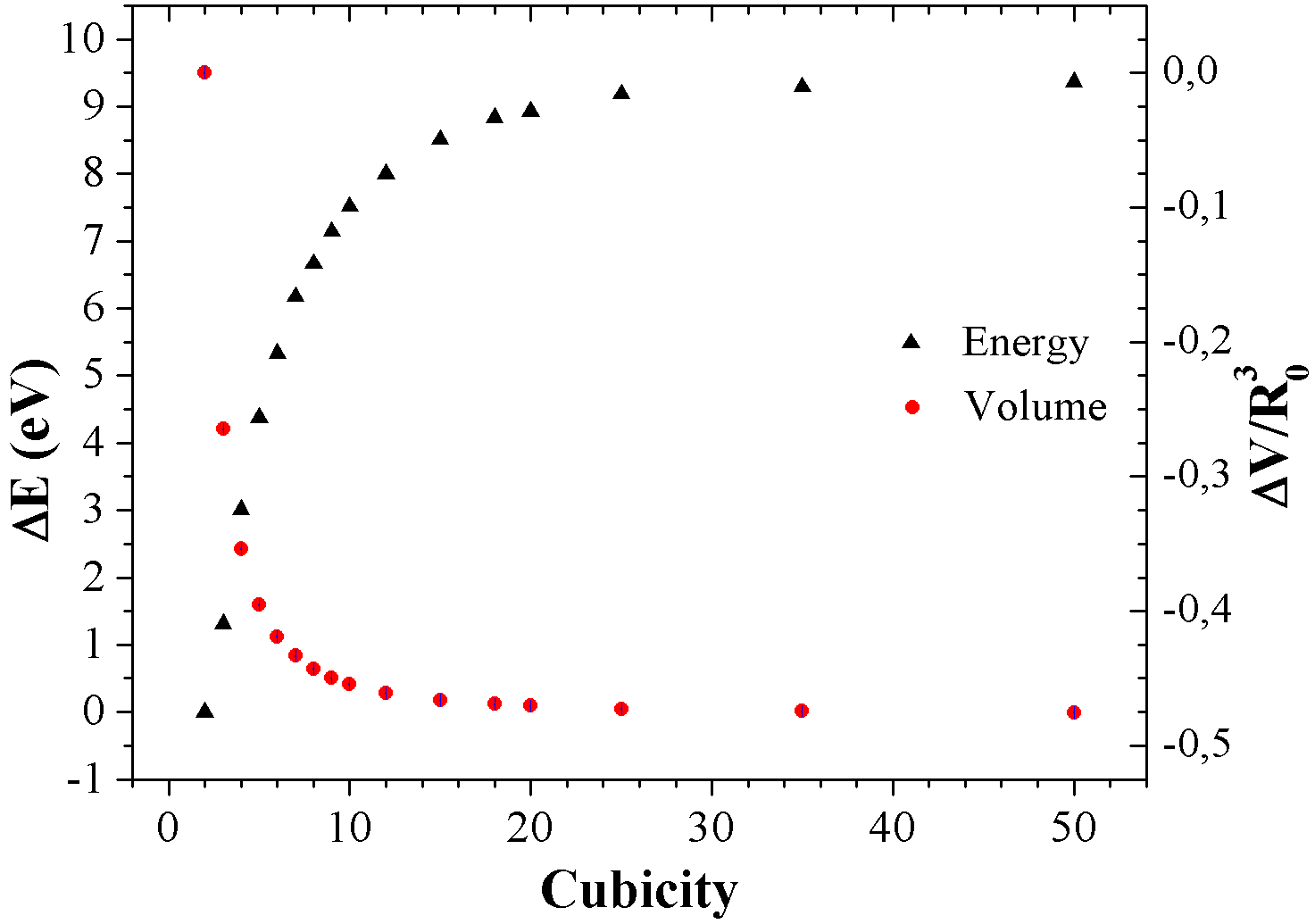}
\caption{Energy cost to deform para-He as function of the cubicity $\eta$ (triangles) and reduction of the lattice volume per atom (circles, normalized to the $P=0$ Van der Waals radius).
\label{fig:1}
}
\end{figure}

Deforming the 1s wavefunction from spherical symmetry to cubic symmetry 
costs energy that we now roughly estimate. We employ Jacobi coordinates for the three-body nucleus (center of mass) and two electron system,
\be
\vec{\pi}_{\rho}=\frac{\vec{p}_{2}-\vec{p}_{1}}{\rho} \qquad \vec{\pi}_{\lambda}=\frac{\vec{p}_{3}-\vec{p}_{1,2}^{\;\,cm}}{\lambda} \qquad \vec{\pi}_{cm}=\vec{p}_{1,2,3}^{\;\,cm}\ .
\ee
As Rayleigh-Ritz variational ansatz we take 
$\psi(\vec{p}_1,\vec{p}_2,\vec{p}_3)\nms\!=\nms\psi_{cm}(\vec{\pi}_{cm}) \psi_{\lambda}(\vec{\pi}_{\lambda})\psi_{\rho}(\vec{\pi}_{\rho})$, with
$\rho$, $\lambda$ being our  parameters to minimize the atomic energy for each family of wavefunctions.

We adopt 
$\psi_{\lambda}\nms\nms=\nms\! e^{-\left(\pi_{\lambda x}^\eta+\pi_{\lambda y}^\eta+\pi_{\lambda z}^\eta\,\right)^{1/\eta}}\nms$ and $\psi_{\rho}\nms\!=\nms\!\! e^{-\left(\pi_{\rho x}^\eta+\pi_{\rho y}^\eta+\pi_{\rho z}^\eta\,\right)^{1/\eta}}$, with $\eta$ a ``cubicity'' parameter.
 For $\eta\!=\!2$ the symmetry of the orbital is spherical. 
For $\eta\!>\!2$ we have a hyperellipsoidal function 
with octahedral symmetry, and for $\eta\to\infty$ (in practice $\eta\sim 10$ is very large already), the shape is perfectly cubic.   

The binding energy is obtained by numerically solving the variational problem for the Hamiltonian of the non-relativistic He atom (neglecting spin and nuclear recoil),
\be
H=-\frac{\hbar^{2}}{2m_{e}}\left(\nabla_{1}^2+\nabla_{2}^2\right)-\frac{Ze^2}{\left|\vec{r}_{1}\right|}-\frac{Ze^2}{\left|\vec{r}_{2}\right|}+\frac{e^2}{\left|\vec{r}_{1}-\vec{r}_{2}\right|}
\ee
$$H\psi=E\psi \qquad \qquad E\leq\frac{\la \psi\ar H\ar \psi\ra}{\la \psi\ar \psi\ra}.$$
The 6 and 9-dimensional integrals resulting are computed with Vegas~\cite{Lepage}, a standard Monte Carlo method.
The computed energy for this interpolating family of functions is shown in figure~\ref{fig:1}. It costs about 9-10 eV to totally deform the Helium atom to cubic symmetry; the accuracy of this computation is about 25\% from the simple-minded variational approach and ignoring the interatomic interaction (appearing only through the pressure in section~\ref{sec:pressure}). For $\eta=2$ the binding of He is underestimated to be 59 eV instead of 79 eV, but since we subtract $E(\eta)-E(2)$, a good part of the error will cancel out in the difference. We feel that higher accuracy at this stage, without experimental data, would be meaningless.

\section{Computation of the pressure required for a given deformation}\label{sec:pressure}

The deformation energy can be provided by Helmholtz's free energy 
 $\Delta E\!=\!P\Delta V$, obtained following the reduction of the total volume of the  \textit{hcp} lattice structure since interstices are reduced. 
The computation of $\Delta V$ is a geometric problem that we defer to the appendix. The outcome has been plotted in figure~\ref{fig:1} as function of the cubicity.
\begin{figure}
\includegraphics[width=0.4\textwidth]{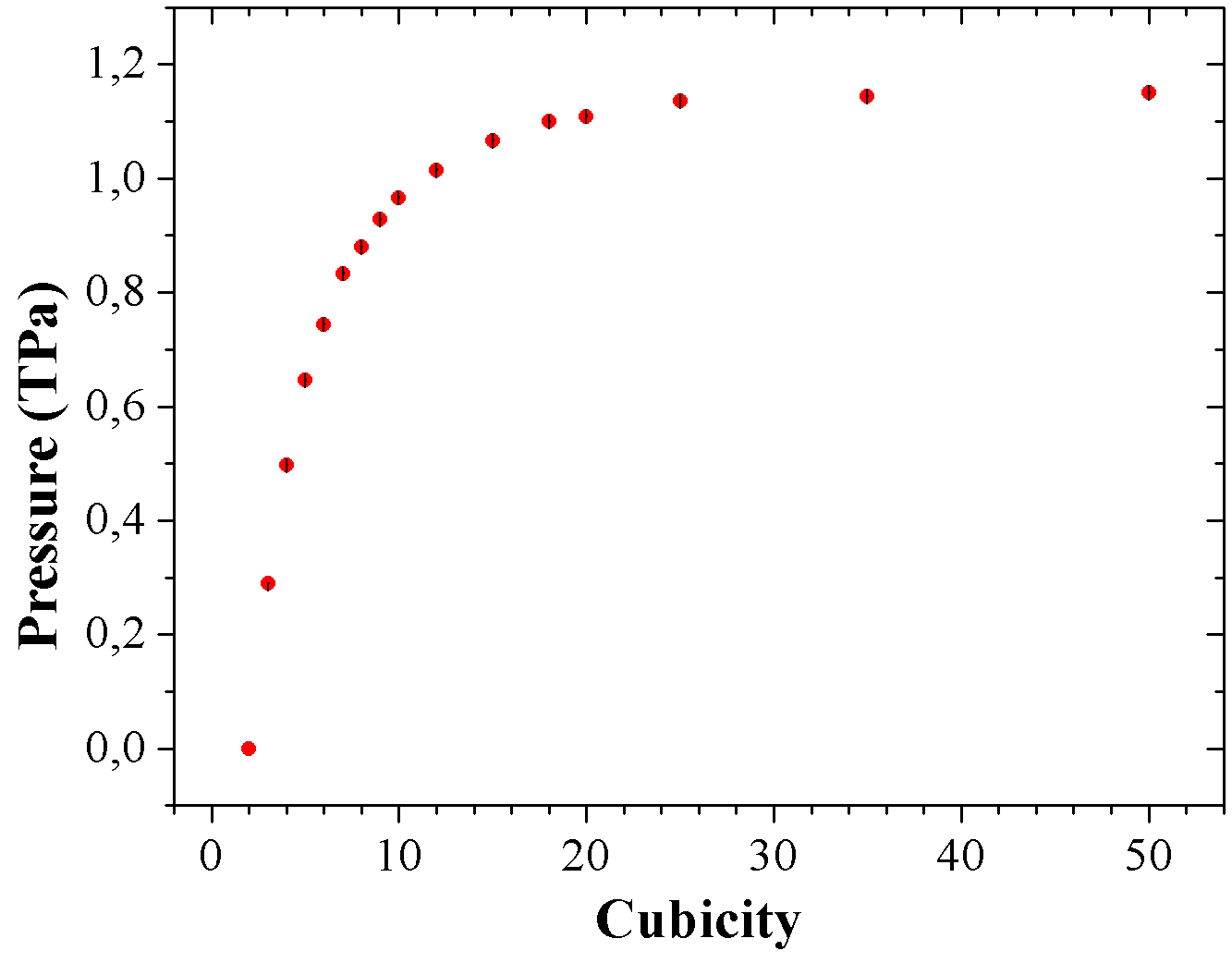} 
\caption{Pressure needed to deform the Helium atom wavefunction from spherical to hyperellipsoidal of given cubicity.
\label{fig:Pofeta}}
\end{figure}

The pressure for which the Helmholtz free energy balances the wavefunction deformation cost is plot, as function of the interpolating cubicity parameter, in figure~\ref{fig:Pofeta}. In the limit of total deformation 
$\eta\rightarrow\infty$, $P=\!1.17\,\text{TPa}$; for 
$P\!=\!500\,\text{GPa}$ the deformation is already significant, with $\eta\!=\!5$. Laboratory studies have reached pressures up to $80\,\text{GPa}$ 
with He and the full $500\,\text{GPa}$ with Hidrogen~\cite{prop_H_He} with the diamond anvil method, so that the wavefunction deformation that we propose, while as yet unseen, is not very far from sight.

Atomic orbital deformation is a concept related to localized electrons around a nucleus. In a metal, delocalized electrons respond to very different physics. So if metallization would occur at a pressure less than what we estimate for cubicity, there would be no hope. 

For He however there are sophisticated density-functional theory computations in the Generalized Gradient Approximation (GGA)~\cite{met_eq.est_he}, 
that suggest that the gap to the conduction band at $T\nms=\!\!0\:\text{K}$
vanishes at  $\rho=17.4\:\text{g/cm}^3$, equivalent to a pressure of
$P\!=17.0\:\text{TPa}$. Alternative simulations based on Diffusion Monte Carlo
(DMC)\cite{montse}, yield an even larger density of $21.3(1)\:\text{g/cm}^3$, corresponding to a pressure  $P\!=25.7\:\text{TPa}$. Both these figures are well above the $O(1)$TPa at which a local, cubic wavefunction in the insulating phase would occur as suggested by our estimates.

\section{Decreased shear modulus}
\begin{figure}[h]
\includegraphics[width=0.4\textwidth]{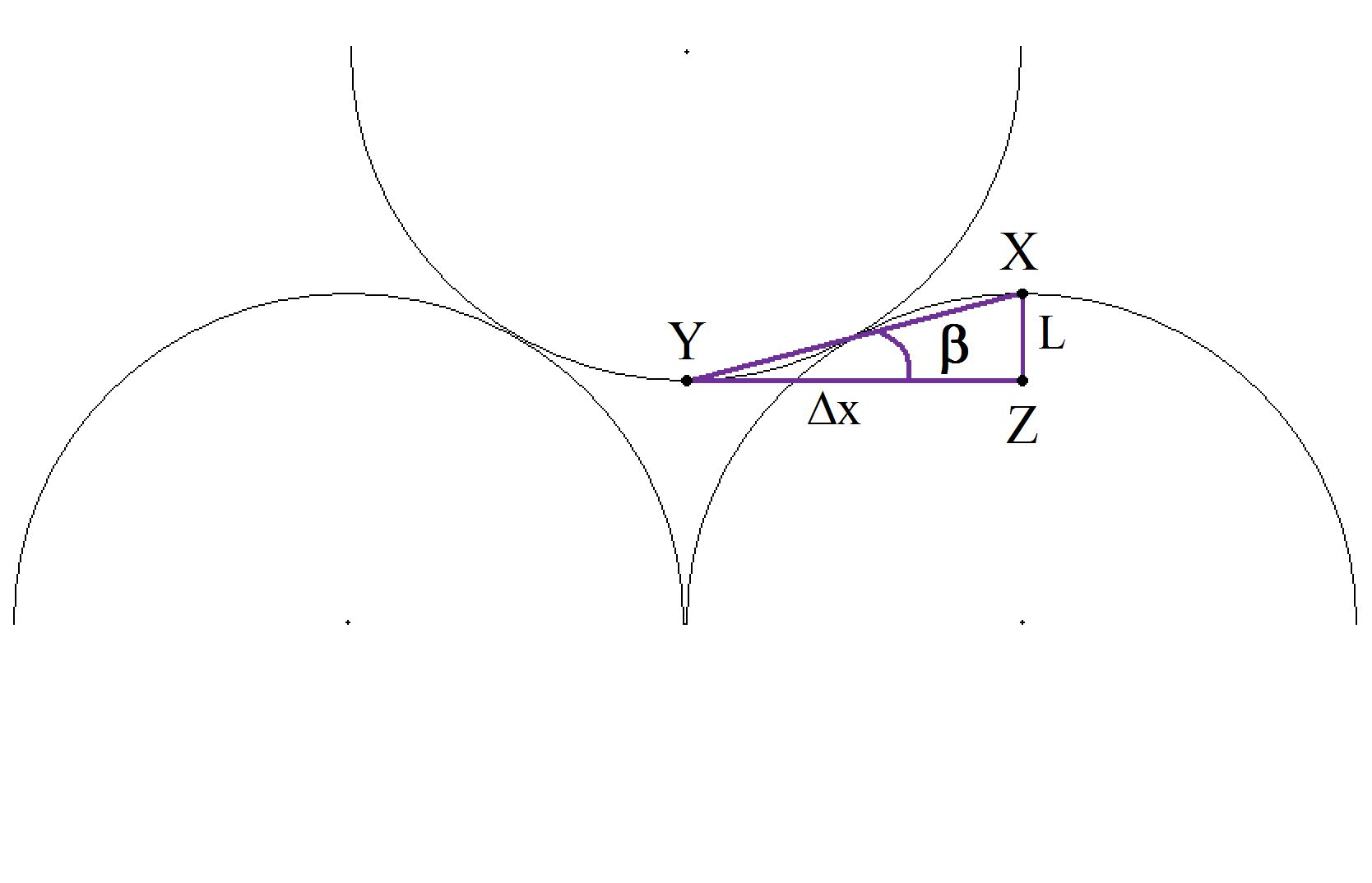} 
\includegraphics[width=0.4\textwidth]{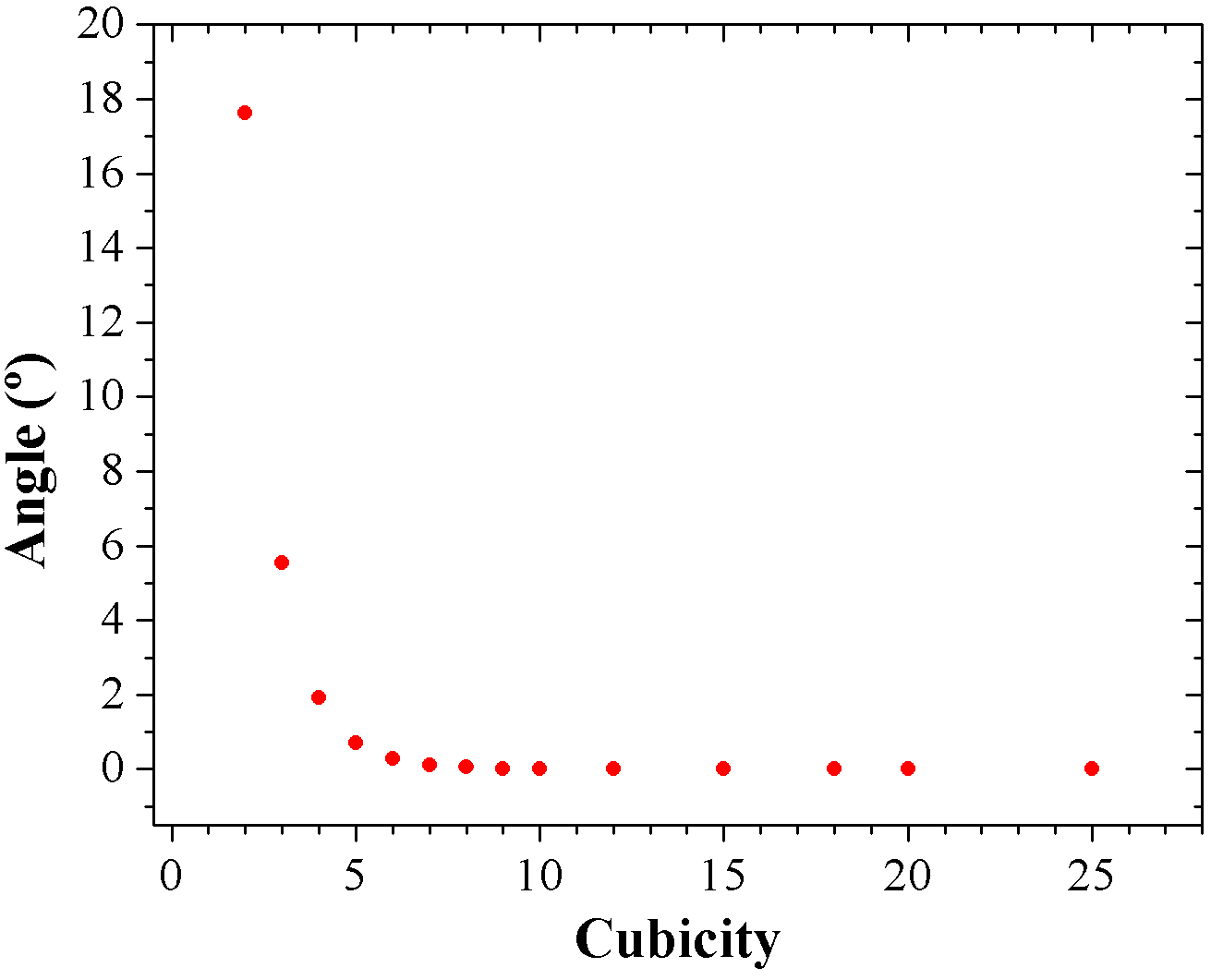}
\caption{Top: Representation of the atomic displacement. 
Bottom: Variation of the displacement angle with cubicity when the crystal undergoes shear stress. \label{fig:angle}
}
\end{figure}

When optimally packing spheres, the successive crystal planes intertwine. A lateral displacement requires (classically) lifting a plane so the spheres can slide (we neglect tunnelling, possible since the atoms are not perfectly rigid bodies, but that opposes Pauli's principle). This lifting requires a force to compensate the pressure. But at large $P$, the atomic deformation allows optimal packing without atoms intruding in neighboring planes (controlled by the function $g(\eta)$ in Eq.~(\ref{gfun})), since the necessary $2R(\eta)$ separation between plane centers is consistent with optimal packing for large $\eta$. Thus, we expect the shear modulus to decrease. A calculation of the complete shear modulus requires also the weak He interatomic potential;  we will refrain from addressing it and only consider differences in the shear modulus between deformed and undeformed He.

\begin{figure}[h]
\includegraphics[width=7.5cm]{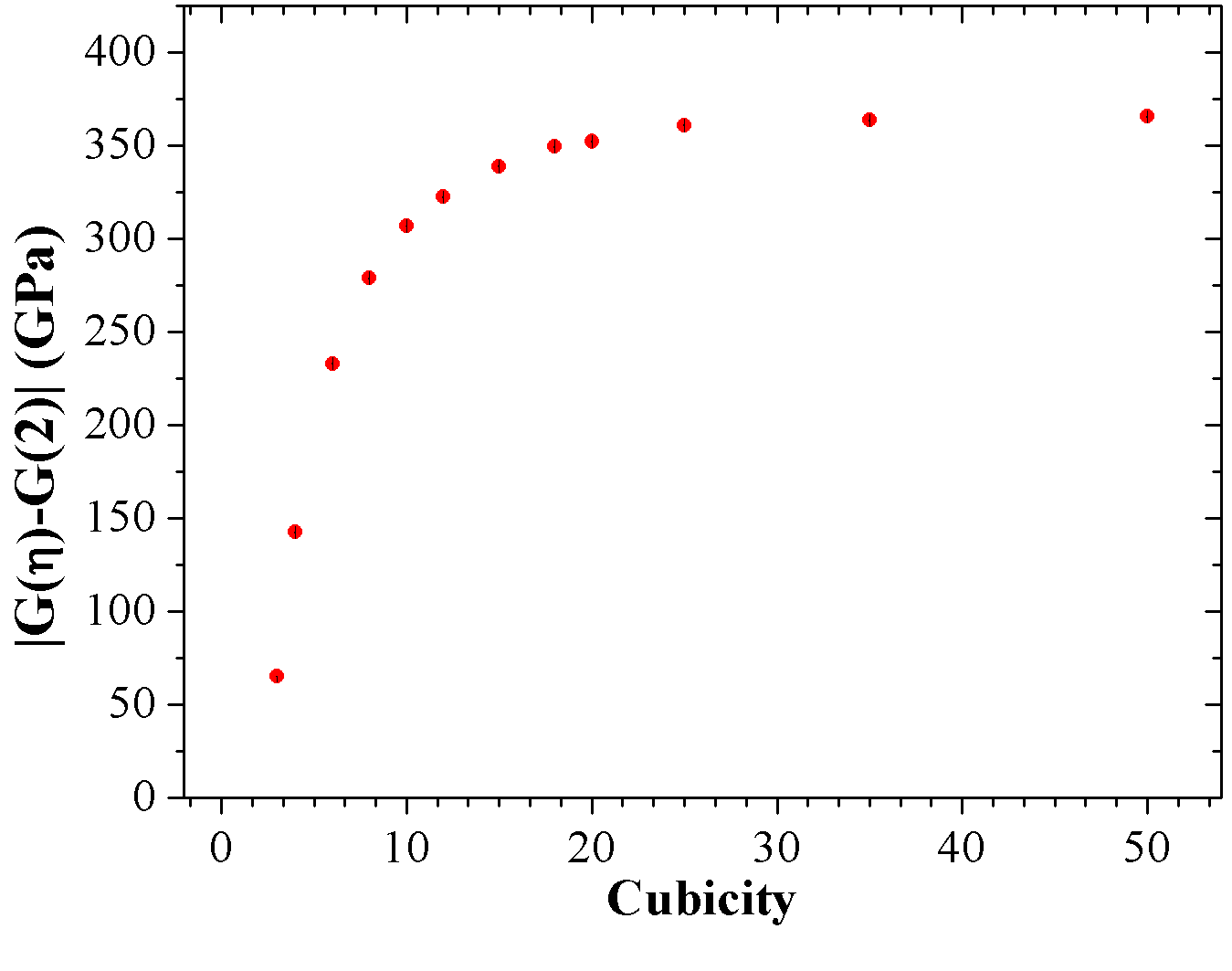}
\includegraphics[width=7.5cm]{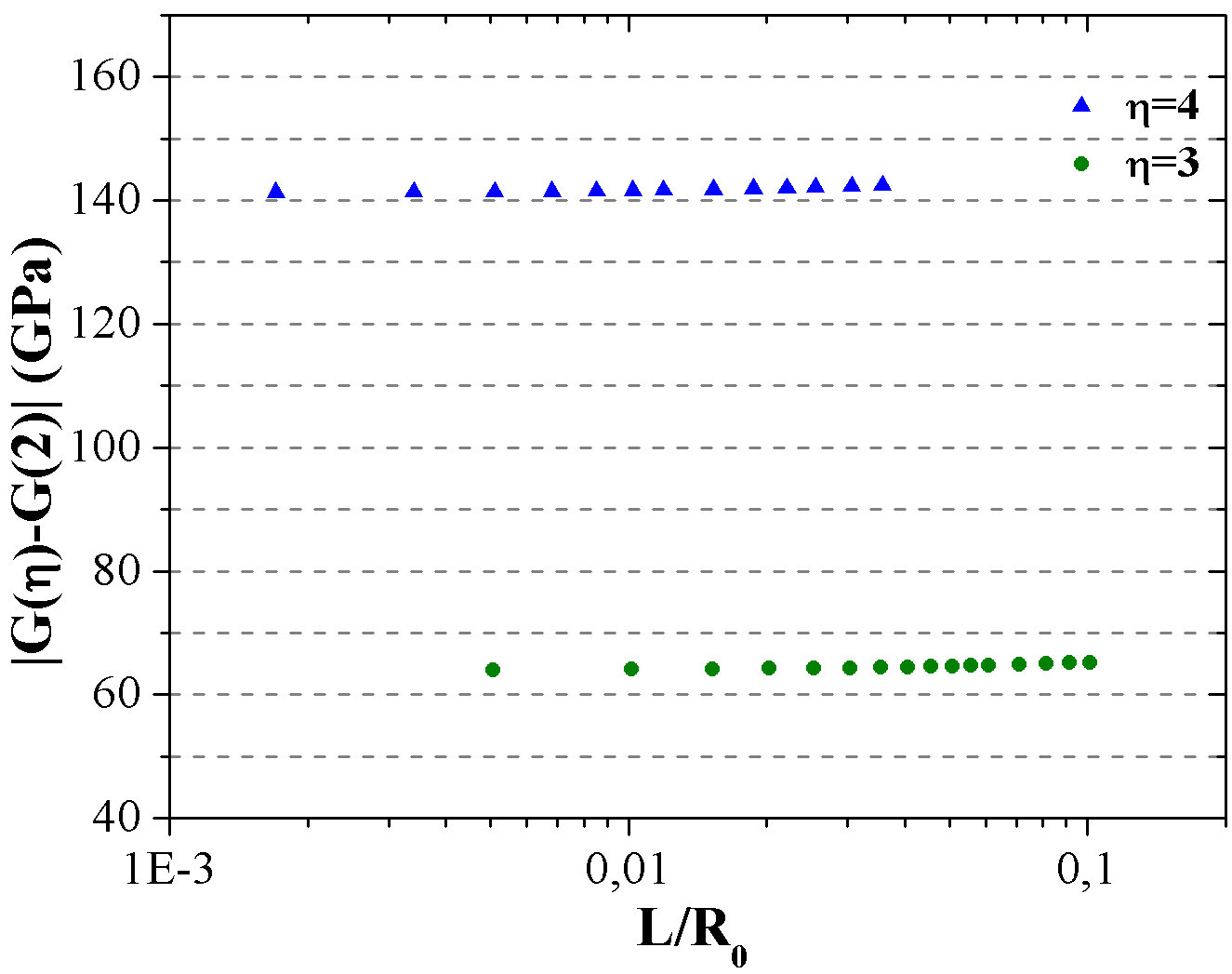}
\caption{Top: Absolute value of the decrease of the shear modulus.
Bottom: Near independence of  $\Delta G$ with the plane-lifting $L$ for two cubicities, as a check of linear elasticity.\label{fig:shear}}
\end{figure}

The shear modulus $G$ is the ratio between stress $\tau=F/A$ and deformation $\ep=x/L$. The displacement parallel to the crystal planes is $x$, and $L$ their lifting, so
\be
G=\frac{\tau}{\ep}=\frac{F/A}{\Delta x/L}=\frac{F/A}{1/\tan\beta}
\label{G1}
\ee
as expressed in terms of  $\beta$, the minimum angle of attack that allows tangential displacement (see figure~\ref{fig:angle}, top panel). This is in turn a function of $\eta$ through the functions $R$, $f$ and $g$ in the appendix. 
Numerically it falls quickly with $\eta$ (figure~\ref{fig:angle}, bottom panel). When it reaches zero the planes can slide sideways without resistance from the pressure, only cohesive forces retain them (taking them into account adds a term to Eq.~(\ref{G1}) and our actual approximation is to take this independent of $\eta$).

The force exerted by the pressure that resists plane lifting
~\footnote{It is of notice that a shear force produces a small crystal dilatation, due to the normal component of the force in the decomposition at the tangent point between atoms, in a similar way to how a body ascends an inclined plane when pulled laterally.}
 is obtained from energy balance, $ F\!\cdot\!L=P\Delta V $, while the volume increase is
\be
\Delta V\!=V(L)\!-\!V(L\!=\!0)=A(L\!+\!d)\!-\nms Ad=A\,L
\label{V_L}
\ee
($d$ is the plane separation in the undeformed structure).
Thus, the contribution of the geometry distortion to $G$ is
\be
\Delta G\!=\!G_{\eta}\nms-G_{2}=P\left[\tan{(\beta_{\eta}(L)})-\tan{(\beta_{2}(L)})\right]
\label{Delta_G}
\ee

The outcome is plotted in figure~\ref{fig:shear}, top plot. As the cubicity $\eta$ increases, the shear modulus $G$ decreases, so $\ar \Delta G\ar$ 
increases to a saturation of about $365\,\text{GPa}$, or about 25\% of the pressure needed to attain that deformation. 
The actual observable prediction is that the shear modulus flattens out at large pressure, instead of growing linearly with it. 

Finally, we provide a check that we remain in the linear elasticity regime in figure~\ref{fig:shear}, bottom panel, where we show that the dependence of
$G$ with the vertical displacement $L$ is very small and only appreciable for small cubicities, and even then it is a correction less than $2\,\text{GPa}$. 

\section{Form factor}

A direct way of ascertaining the structure of a charge distribution is to probe it by scattering an electron or photon beam. Shape information about the target is encoded in the elastic form factor. This is defined so that the charge density of the object under study is normalized to 1,
$\int d^3x \rho(x)=1$. The charge density is easily computed from the variational wavefunctions $\psi_\lambda$ and $\psi_\rho$ as in section ~\ref{deformedenergy} (do not confuse this charge density with Jacobi's coordinate for the three body problem). Since we aim to calculate the form factor of an atom, the nonrelativistic approximation is sufficient, so that
\be
F(\vec{q}\;^2) = \int \frac{d^3 q}{(2\pi)^3} e^{-i {\vec{q}}\cdot {\vec{x}}} \rho(x) 
\ee
with $F(0)=1$ due to the charge normalization of the two-electron cloud (overall neutrality is guaranteed by the He nucleus, which, being approximately pointlike, provides a trivial constant contribution that we understand as subtracted). 

We carry out this integration numerically. We imagine that the compressed atomic  system (whose deformation we should like to measure) is analyzed with X-ray photons of momentum 
$\vec{q}$.
Such X-ray diffraction experiments have been carried out as already discussed~\cite{xray_He}, though at smaller pressures as we would need here, but they might be performed in the future.
 
We now briefly observe what simple changes of the form factor one would hope to find due to the atomic deformation.

\begin{figure}
\includegraphics[width=0.4\textwidth]{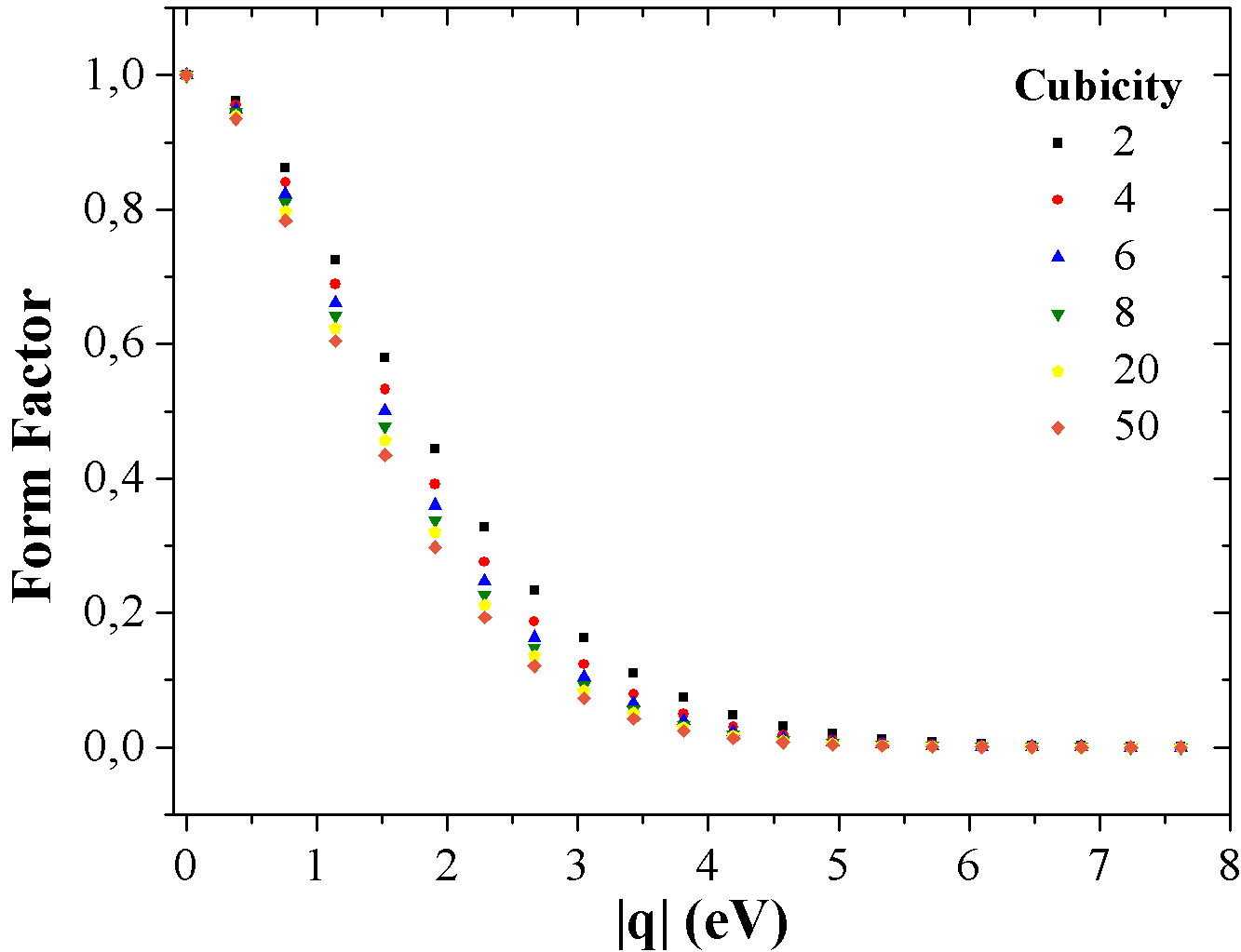}
\caption{Form factor for different cubicities (grade of the hyperellipsoidal wavefunction) 
as function of the modulus of the incident photon's momentum. \label{fig:var_N_qpol0_qaz0}
}
\end{figure}

The numeric computation of the form factor as function of $\|\vec{q}\|$ is given in figure \ref{fig:var_N_qpol0_qaz0}. Since our wavefunction ansatz is a simple exponential, the Fourier transform yields a falling rational form factor as seen in the figure. The difference in the momentum dependence to the spherical case is subtle and likely not isolatable from experimental data with finite error bars.  We observe that the form factor decreases a bit quicker for higher deformation.

\begin{figure}
\includegraphics[width=0.45\textwidth]{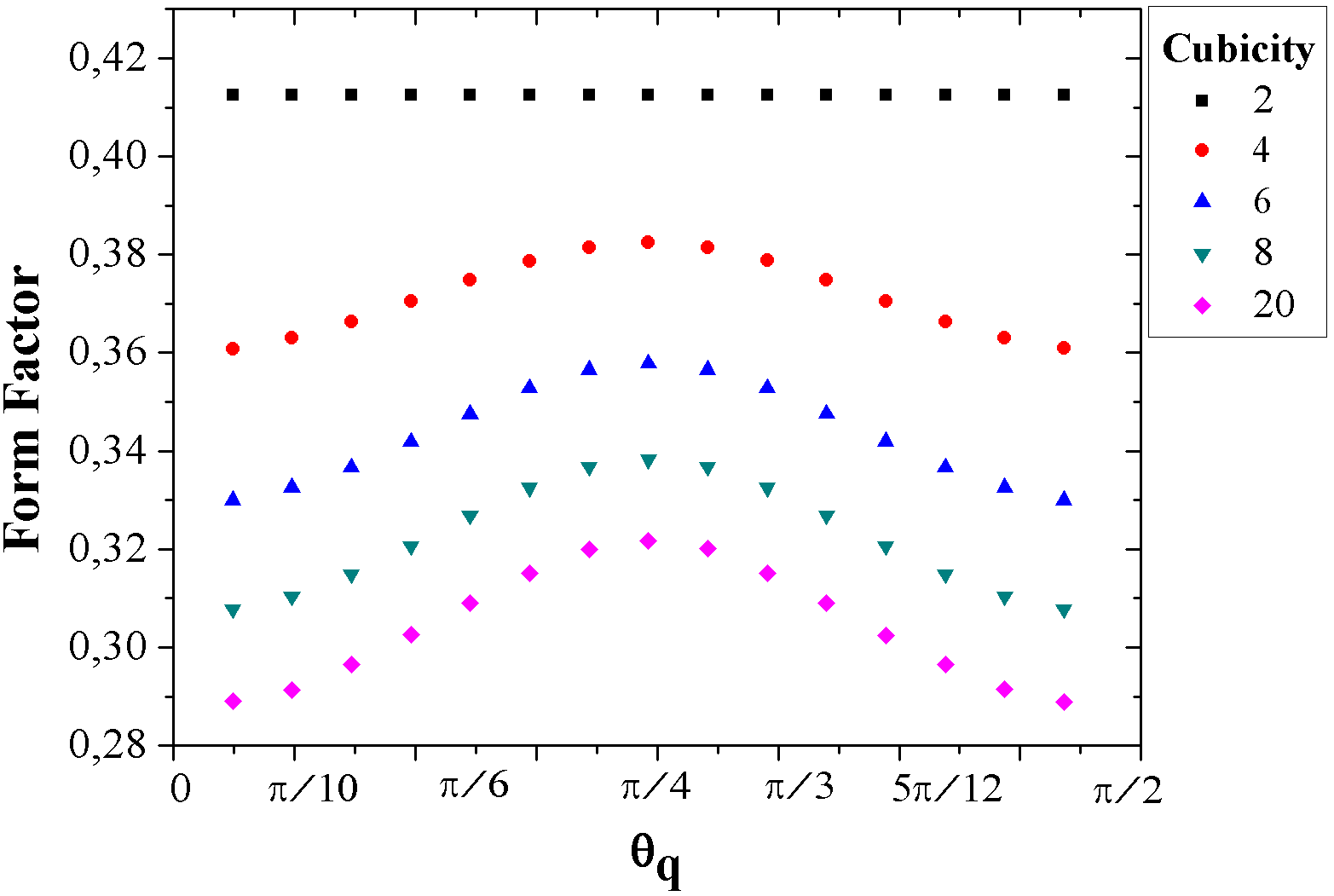}
\includegraphics[width=0.4\textwidth]{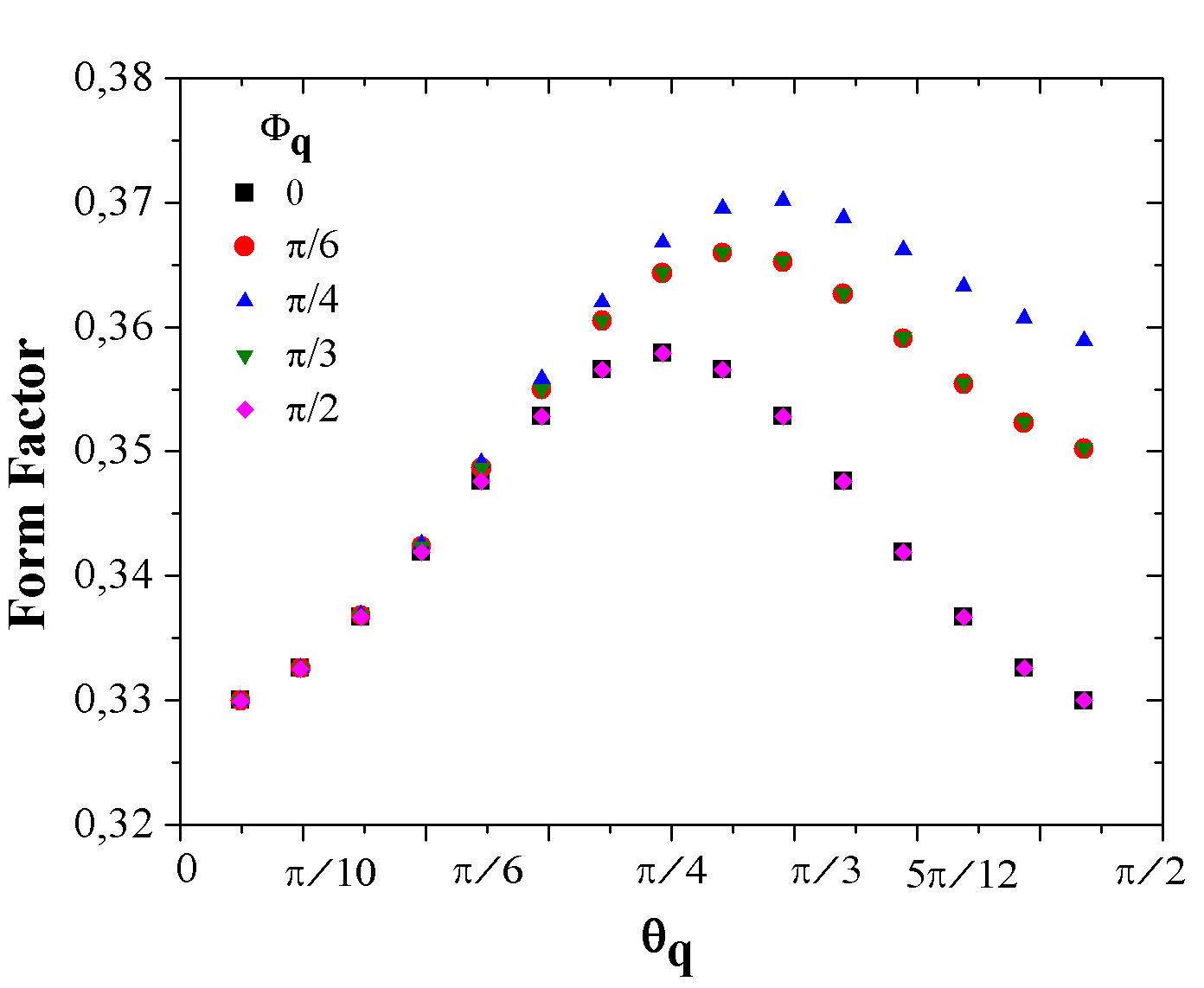}
\caption{Top: Form factor as function of $\theta_q$ for selected cubicities, 
with fixed $\|\vec{q}\|\!=\!4\;\text{eV}$ and $\phi_q\!=\!\pi/2$. 
Bottom: Variation with the polar angle $\theta_q$ of the form factor for fixed $\|\vec{q}\|\!=\!2\;\text{eV}$ and $\eta=6$ showing the dependence on the azimuthal angle. 
Note that pairs of numerical data sets fall on top of each other due to
the $\pi/2$ rotational symmetry of the hyperellipsoidal wavefunctions.\label{fig:var_qpol}
}
\end{figure}

More interesting is the breakup of azimuthal symmetry by the cubic deformation.
We give the change of the form factor with that atomic deformation as function of the polar scattering angle in figure \ref{fig:var_qpol} (top panel). 
The form factor changes significantly with varying cubicity; for $\eta=2$ the form factor is independent of the measurement angle because the charge distribution is spherically symmetric, property maintained by the Fourier transform (top squares following a horizontal line) while for deformed wavefunctions the form factor is not only smaller but also acquires a remarkable angular dependence which is the tell-tale signature. 

There is of course also a dependence on the azimuthal angle  that we show in figure~\ref{fig:var_qpol} (bottom panel) for a $\|\vec{q}\|$ fixed and varying polar angle.
The reduced rotational symmetry of the deformed atoms for a fixed cubicity parameter
causes some degeneracies that provide a practical check of the computer code.

\section{Outreach and discussion}
In section~\ref{deformedenergy} we employed a direct variational calculation,
but there is another way of thinking of a wavefunction with cubic symmetry, in terms of a spherical-harmonic decomposition:  a linear combination of the 1s orbital (with weight $\sim 74$\%), a 3d orbital (about 25\%) and residual amounts of higher $g$, $i$, etc. orbitals. The energy estimated this way is similar to what we have obtained in section~\ref{deformedenergy}.

Extreme pressures are reached not only in the laboratory but also in compact stars. In the cooling of white dwarves, the metal-insulator transition of He changes the carriers of energy from $e^-$ to $\gamma$~\cite{met_eq.est_he}. Since metallization occurs deep in the star~\cite{met_eq.est_he,montse}, there is ample room for atomic deformation in the crust, and as we suggest it affects the mechanical properties of the star.

\begin{figure}
\includegraphics[width=0.4\textwidth]{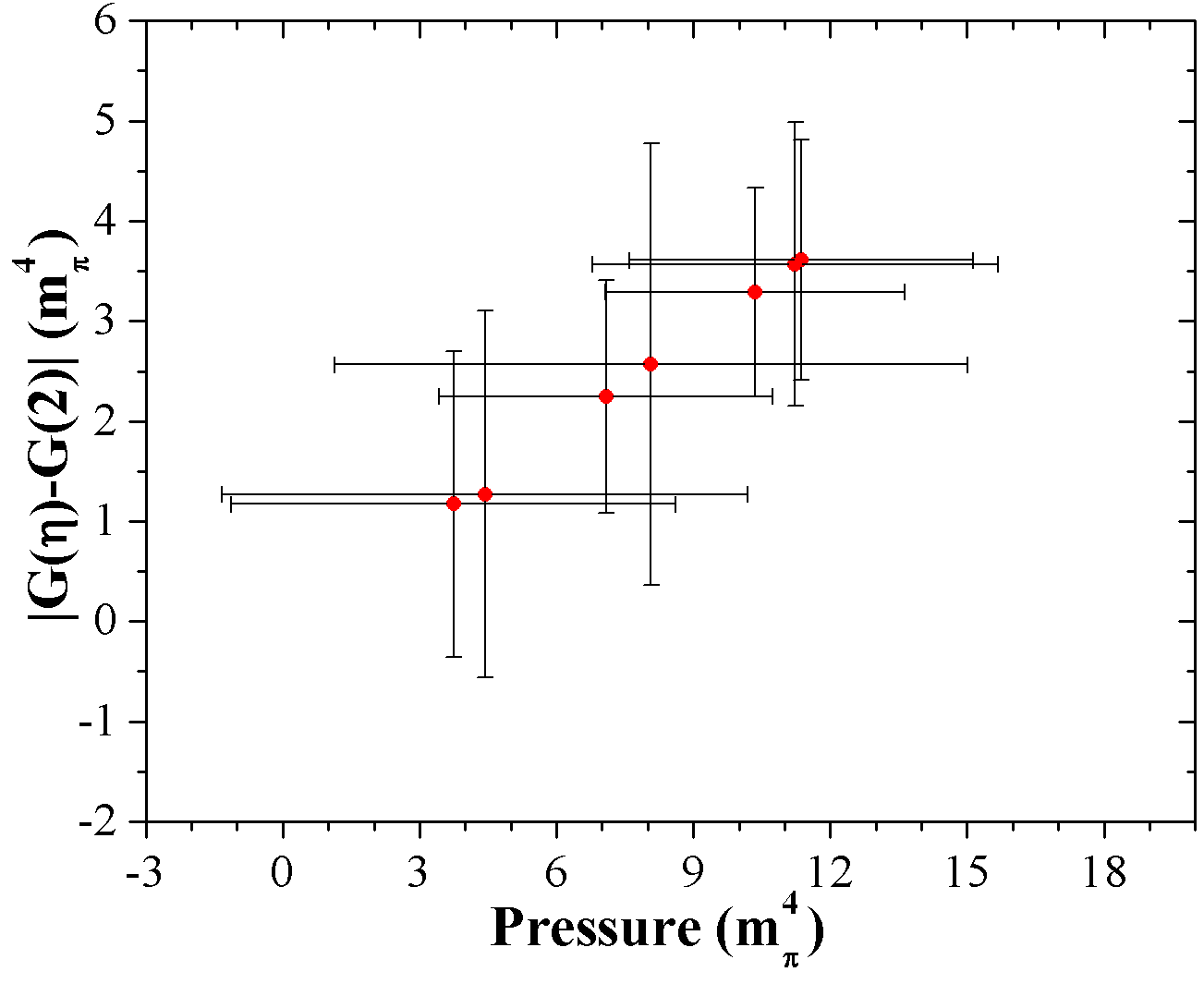}
\caption{Absolute value of the decrease of the shear modulus as a consequence of neutron deformation in a neutron star, were an anisotropic crystal to develop, as function of the pressure. The cubicity is in the interval $(2,18)$. \label{fig:nstar}
}
\end{figure}

Neutrons in neutron stars are also subject to extreme pressures, and though often treated as a superfluid, they may crystallize~\cite{canuto,haensel}.
Heisenberg's uncertainty principle
$\Delta x\Delta p\simeq1\rightarrow\Delta E=\frac{\Delta p^2}{2m_{n}}\simeq\frac{1}{2m_{n}\Delta x^2} $ suggests the pressure necessary to confine the neutron in a lattice cell, with the volume gain estimated from the star's density and Kepler's packing fraction~\cite{pack}, 
$\Delta V\!=\left(f_{\rm crystal}-f_{\rm fluid}\right)V_{cell}$ with $V_{\rm cell}\sim 2R_{n}^3$, $R_n\!=0.8775\:\text{fm}$. 
In appropriate units ($m_\pi=138$ MeV), the pressure is $P\simeq\!1.5\,\text{m}_{\pi}^4$, which occurs~\cite{felipe_dobado} at a depth of 3 km below the star surface (typical radii are 10 km).

In such conditions, neutron deformation  is a possibility, and we calculated in~\cite{felipe} that the neutron mass as a 3-quark system increases by $150\;\text{MeV}$ between $\eta\!=\!2$ and $\eta\!=\!18$.
The necessary pressure saturates at about $\sim\!11\,\text{m}_{\pi}^4$, and the change of the shear modulus as function of the pressure is shown in figure~\ref{fig:nstar};
this may be of interest to compute the sismic response of neeutron stars~\cite{Watts}.

To conclude, we feel that wavefunction deformation is an interesting qualitative phenomenon possible in highly compressed systems, with applications beyond laboratory experiments to astrophysics, and hope that it is born in future, more sophisticated calculations, as well as in experimental work.

\appendix
\section{Packing of hyperellipsoids}
\begin{figure}[h]
\includegraphics[width=0.4\textwidth]{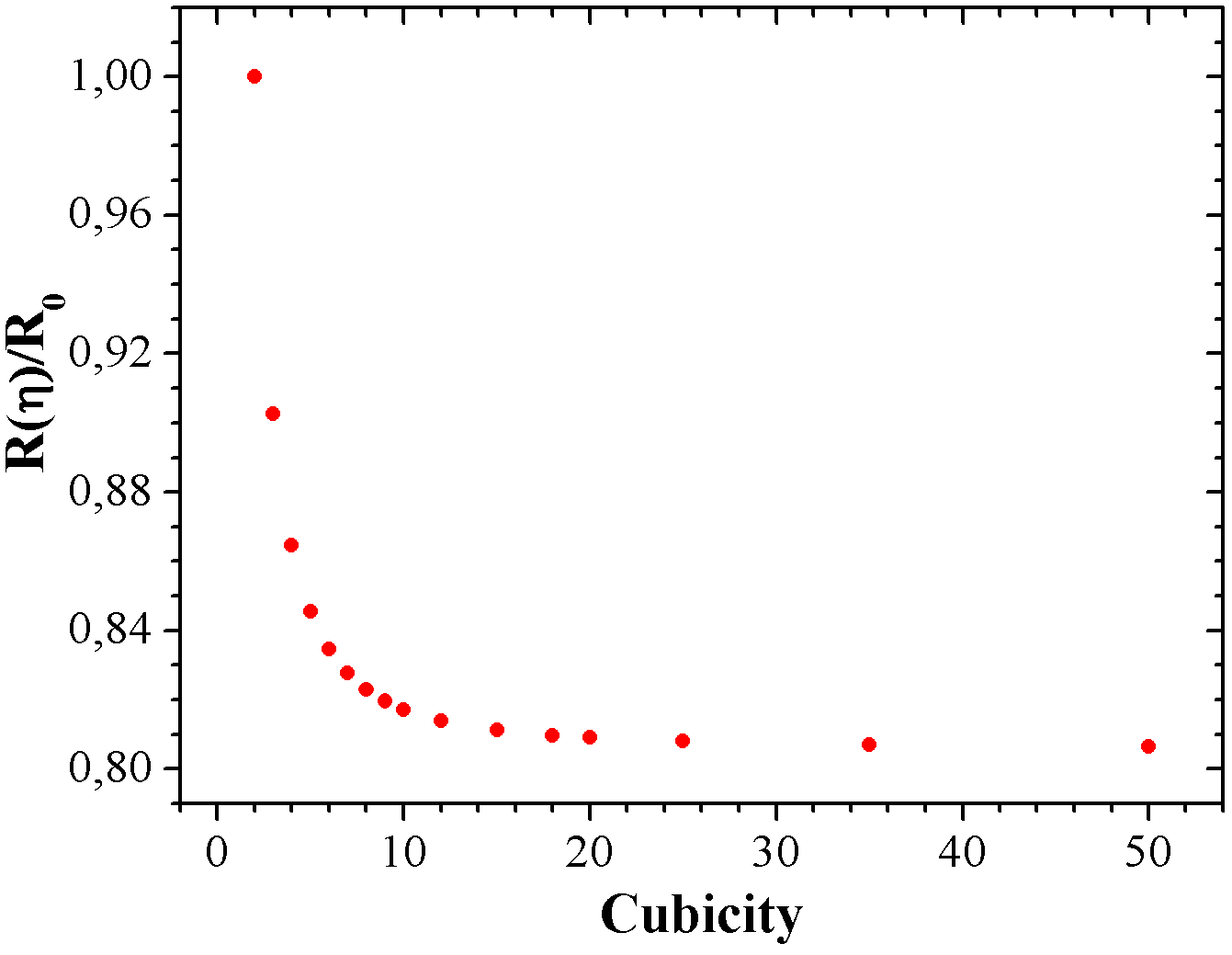}
\centering
\caption{To keep the hyperellipsoid's volume constant notwithstanding the deformation, its (short) radius decreases as shown.
\label{fig:ratio}}
\end{figure}
\begin{figure}[h]
\includegraphics[width=0.3\textwidth]{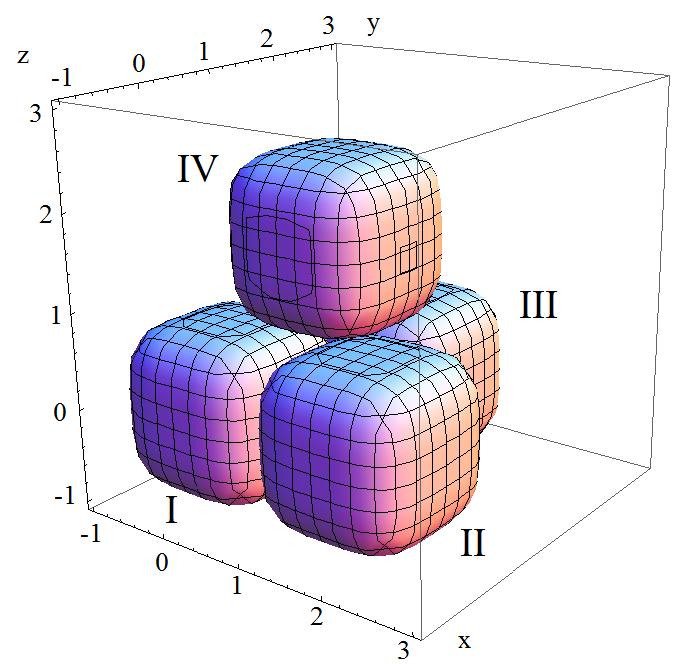}
\centering
\caption{Atomic disposition in a unit cell at close-packing.
\label{hiperelipsoides_tetraedro}}
\end{figure}

Hyperellipsoidal coordinates, $\phi\!\in\![0 , 2\pi]$, $\theta\!\in\![0 , \pi]$, with
$x=a_1 \cos^{2/\eta}\!\theta \cos^{2/\eta}\!\phi$,
$y=a_2 \cos^{2/\eta}\!\theta \sin^{2/\eta}\!\phi$, and
$z=a_3 \sin^{2/\eta}\!\theta$ are used for the hyperellipsoid's volume.

First we obtain the area of the hyperellipse obtained upon cutting the figure with a horizontal plane by means of Green's theorem $ A\!=\!\frac{1}{2}\oint_{C}\,(x\,dy-y\,dx)$, which is $A=\frac{4ab}{\eta}\:B\!\left(\frac{1}{\eta},\frac{1}{\eta}+1\right)$ in terms of Euler's beta function. Letting now the axes of the hyperellipse vary with height or polar angle, $a(\theta)$, $b(\theta)$, 
we obtain the known formula for the volume~\cite{Dordrecht},
\be \label{volumen}
V(\eta)=8R(\eta)^3\;\frac{\displaystyle \left[\Gamma\!\left(1+\frac{1}{\eta}\right)\right]^3}{\displaystyle \Gamma\!\left(1+\frac{3}{\eta}\right)}
 \ .
\ee
If we understand the radius of the hyperellipsoid as the minimum distance from its surface to its symmetry center, the relation between this radius and that of the sphere of equal volume is, because of Eq.~(\ref{volumen}), 
\be
R(\eta) = \left[\frac{\pi}{6}\right]^{1/3}\frac{\displaystyle \left[\Gamma\!\left(1+\frac{3}{\eta}\right)\right]^{1/3}}{\displaystyle \Gamma\!\left(1+\frac{1}{\eta}\right)}\: R_0 \ .
\label{ratio}
\ee
The resulting $R(\eta)$ is pictured in figure~(\ref{fig:ratio}).
Obviously for large cubicity, $\lim_{\eta\to\infty}R(\eta)=\left[\pi/6\,\right]^{1/3}R_0$ because the cube's volume is $8R(\eta\to\infty)^3$.

We are ready to compute the packing fraction; consider the first octant of a given hyperellipsoid I and the adjacent ones in a close-packing structure that are tangent in that octant, numbered II through IV in figure~\ref{hiperelipsoides_tetraedro}.

When $\eta\nms=\nms2$, the sphere's centers are $(0,\,0,\,0)$, $(2R_0,\,0,\,0)$, $(R_0,\,\sqrt{3}R_0,\,0)$ and $(R_0,\,\sqrt{1/3}R_0,\,\sqrt{8/3}R_0)$.
Upon deforming the figures, they may move closer since space is used more efficiently, and the centers become $(0,\,0,\,0)$, $(2R(\eta),\,0,\,0)$,
$(R(\eta),\,f(\eta)R(\eta),\,0)$ and $(x,\,y,\,g(\eta)R(\eta))$, where the auxiliary functions $R(\eta)$, (determined numerically), $x$, $y$, $f(\eta)=(2^{\eta}-1)^{1/\eta}$  and $g(\eta)$
\be \label{gfun}
g(\eta)=\left[2^{\eta}-1-\left(\frac{f(\eta)^2-1}{2f(\eta)}\right)^{\eta}\right]^{1/\eta}
\ee
are found requesting that the four hyperellipsoids 
\ba
	x^\eta+y^\eta+z^\eta=R(\eta)^\eta
	\label{I} \\
	\left(x-2R(\eta)\right)^\eta+y^\eta+z^\eta=R(\eta)^\eta
	\label{II} \\
	\left(x-R(\eta)\right)^\eta+\left(y-f(\eta)R(\eta)\right)^\eta+z^\eta=R(\eta)^\eta
    \label{III} \\
	\left(x-R(\eta)\right)^\eta+\left(y-\frac{f(\eta)^2-1}{2f(\eta)}R(\eta)\right)^\eta+ \nonumber \\
\left(z-g(\eta)R(\eta)\right)^\eta=R(\eta)^\eta
    \label{IV}
\ea
are tangent at precisely one point.

\begin{figure}[b]
\centering
\vspace{4mm}
\includegraphics[width=0.4\textwidth]{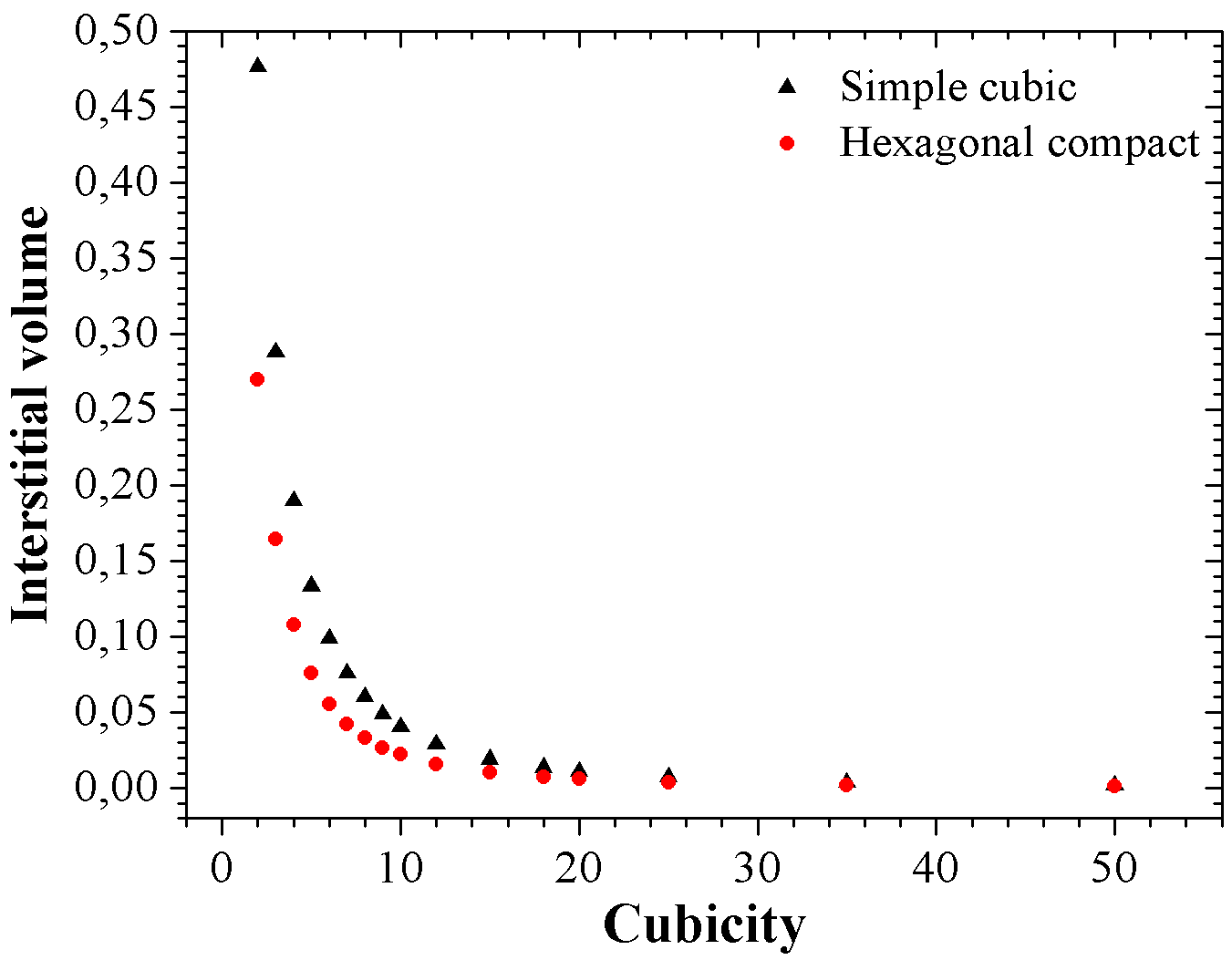}
\caption{\label{vol_interst}
Interstitial volume in the \textit{hcp} and \textit{cs} structures.
The volume gained upon occupation of the interstices does not differ so much between the two packings.}
\end{figure}

Once we have the size and position of the deformed figures we can calculate the occupied volume fraction. Without loss of generality we may choose 
$R_0\!=\!1$ and take as total volume a cube of radius 
$R(\eta)$ with a vertex on the origin, that just leaves hyperellipsoid II out; but III and IV, because of the closely packed structure, fill some of the interstitial space.
 The octant of hyperellipsoid I occupies $V_I=\frac{\pi}{6}$. The other two occupy
\ba
V_\text{III}=
	& &\int_{R-\left(R^{\eta}-\left|R-fR\right|^{\eta}\right)^{1/\eta}}^{R}
	\int_{fR-\left(R^{\eta}-\left|x-R\right|^{\eta}\right)^{1/\eta}}^{R}
\nonumber \\ 
	& & \int_{0}^{\left(R^{\eta}-\left|x-R\right|^{\eta}-\left|y-fR\right|^{\eta}\right)^{1/\eta}}
	\nts\nts\nts\nts dxdydz
\label{V_III} 
\ea
\ba
V_\text{IV}=
	& & \int_{R-\left(R^{\eta}-\left|R-gR\right|^{\eta}\right)^{1/\eta}}^{R} 	\int_{y_{inf}}^{y_{sup}} \\ \nonumber
   & & \int_{gR-\left(R^\eta-\left|x-R\right|^\eta-\left|y-\frac{f^2-1}{2f}R\right|^\eta\right)^{1/\eta}}^{R}
    dxdydz \ .
\label{V_IV}
\ea
The resulting occupied fraction can be subtracted from 1 to obtain the wasted or interstitial fraction depicted in figure~\ref{vol_interst}.
For $\eta=2$ we reproduce Kepler's result for optimal packing, about 26\% inefficient~\cite{pack}. As $\eta$ grows we see that this number decreases and eventually vanishes (the packing of cubes leaves no interstitial space). For comparison we also show the rather trivial result for a simple-cubic lattice. For qualitative reasoning, we see that the actual type of lattice is not very important.

\begin{acknowledgments}\noindent
FLE supported by Spanish grant FPA2011-27853-C02-01 and  CPAN - Proyecto Consolider-Ingenio 2010; 
PCP by grants FPA 2013-41267 and FPA 2010-17142.
\end{acknowledgments}


\end{document}